\documentclass[11pt]{article}
\usepackage{graphicx}
\usepackage{latexsym}
\usepackage{amsmath,amssymb}

\usepackage{epsfig}
\usepackage{graphicx}

\usepackage{graphicx}
\usepackage{amssymb,amsmath,times}

\def\be{\begin{equation}}
\def\ee{\end{equation}}
\def\ba{\begin{eqnarray}}
\def\ea{\end{eqnarray}}
\def\la{\langle}
\def\ra{\rangle}
\def\a{\alpha}
\def\b{\beta}

\def\h{\hskip 1cm}

\def\A1{A_{-1}}
\usepackage{epsfig}
\usepackage{latexsym}
\usepackage{amsmath,amssymb}
\begin{document}
%%\begin{titlepage}
%\vspace{4cm}
\begin{center}
{\large{ \textbf{Quantum information reclaiming after amplitude damping
}}}\\
\vspace{1cm}
Laleh Memarzadeh  \footnote{
email: laleh.memarzadeh@unicam.it},\h Carlo Cafaro\footnote{email: carlo.cafaro@unicam.it}, \h
Stefano Mancini\footnote{email: stefano.mancini@unicam.it
}\\
\vspace{1.5 cm}
{School of Science and Technology, University of Camerino
, I-62032 Camerino, Italy}
\vspace{1cm}
\end{center}
%\maketitle
%\affiliation{Dipartimento di Fisica, Universit\`{a} di Camerino, I-62032 Camerino, Italy}
\begin{abstract}
We investigate the quantum information reclaim from the environment after amplitude damping has occurred. In particular we address the
question of optimal measurement on the environment to perform
the best possible correction on two and three dimensional quantum systems.
Depending on the dimension we show that the
entanglement fidelity (the measure quantifying the correction performance) is or is not the same for all possible
measurements and uncover the optimal measurement leading to the maximum entanglement fidelity.\\
\\
PACS number: 03.67.Pp, 03.65.Yz
\end{abstract}

\vspace{1 cm}
\section{Introduction}\label{intro}
Decoherence is a fundamental problem for the realization of quantum information tasks.
The unavoidable interaction of a system with the surrounding environment
disturbs the state and causes decoherence \cite{Zurek}. This situation is described by
a unitary interaction between the system and the environment. The state
change of the system can be described by a
CPT map (completely positive trace preserving map). Usually this implies an
irreversible flow of information from the
system to the environment with a consequent wash out of coherence features of the system. Having
access to the environment one could think to reclaim the information lost into the environment and
restore the (quantum) coherence features of the system with a proper action on it.
This would be a typical example of feedback (closed loop) control [2] in which the actuation
on the system would be based on classical information gathered from environment.
This method is practical for
systems interacting with sufficiently controllable environments, that is, environments on which
 the necessary measurements can be performed (e.g. in experiments
inside cavities).\\
\\
Recently, a lot of attention has been
devoted to the feedback control scheme from different aspects.
In \cite{HaydenKing,SVW} the capacity for this scenario has been studied and in
\cite{BCD} it has been shown that in the limit of infinite number of iteration of a map
on qubits and qutrits, the effect of
decoherence can completely be removed.
In \cite{QLF} it has been shown that if the initial state of the environment is pure, then any measurement on the environment corresponds to a specific Kraus representation of the
map describing the evolution of the system. For a given measurement the best recovery scheme has also been introduced \cite{QLF}. However, it is clear that the success of this technique resides on the optimality of
measurement.\\
\\
Here we address the question of what kind of measurement can provide the highest achievable coherence
after performing the optimal recovery. In particular we consider the nontrivial case of amplitude
damping channel describing the loss of energy (hence information) from system to environment.
Due to the correspondence between measurement and Kraus representation of the map, any classical
result of measurement $\alpha$ can be interpreted as the error $t_{\alpha}$ (Kraus operator) occurred on the system.
Since the canonical Kraus operators are independent of
each other, a naive guess about the optimal measurement might be the one corresponding to canonical
Kraus representation.
Actually we show that for qubits all possible measurements on the environment are
equivalent in the sense of recovering the errors by feedback control. We also show that
for amplitude damping channels acting on qutrits the situation is different and the measurement
corresponding to canonical Kraus representation is not the optimal measurement on the environment. \\
\\
The layout of the paper is as follows. In Section \ref{QF}, we briefly present the main conceptual and computational tools
needed to restore quantum coherence by means of a quantum feedback control scheme. We introduce
amplitude damping channel in Section \ref{AD-Channel}. For $d=2$, we show that
all decompositions or all measurement on the environment are equivalent for restoring quantum coherence via feedback control. In Section \ref{AD-3}, we study
the amplitude damping channel for qutrits. We find that for $d=3$ the canonical Kraus decomposition is sub-optimal.
Indeed, by means of analytical methods, we construct a class of (non-canonical) Kraus decompositions leading
to an higher performance than the one obtained via the canonical decomposition. The paper concludes in
Section \ref{diss} with a discussion.
\section{Quantum feedback control scheme}\label{QF}
In this Section we describe the general scheme for error correction using the information
we gain by measuring the state of environment. The evolution of state in interaction with
an environment or the action of any quantum channel on the input state can be described
by considering the unitary evolution of the system together with the environment. Considering
Hilbert spaces of initial and final states
$\mathcal{H}_1$ and  $\mathcal{H}_2$ and also the Hilbert spaces for initial and final environments $\mathcal{K}_1$ and $\mathcal{K}_2$,  the
general evolution can be described by a unitary operator $U: \mathcal{H}_1\otimes \mathcal{K}_1\rightarrow \mathcal{H}_2\otimes \mathcal{K}_2$. The final state of the system is given by tracing over the environment after the interaction,
\begin{equation}
T(\rho)=Tr_{_{\mathcal{K}_2}}[U(\rho\otimes\sigma) U^{\dag}],
\end{equation}
where $\rho$ and $\sigma$ are positive trace class operators belonging to $\mathcal{L}(\mathcal{H})$, the space
of all linear operators
on Hilbert space $\mathcal{H}$. The map $T:\mathcal{L}(\mathcal{H}_1)\rightarrow \mathcal{L}(\mathcal{H}_2)$ is a CPT map
which describes the channel or evolution of a system after interaction with environment.
Our aim is to gather information about the errors occurred on the system by measuring the environment after interaction with the system . A general POVM measurement on $\mathcal{K}_2$ is described by $M_{\alpha}\in \mathcal{L}(\mathcal{K}_2)$ where
\begin{equation}\label{measurment}
\sum_{\alpha}M_{\alpha}=I, \hskip 1 cm M_{\alpha}>0.
\end{equation}
The index $\alpha$ labels the classical result of measurement. To understand how this classical information
can be used to gain information about the final state of the system, let us consider
an arbitrary observable $A\in \mathcal{L}(\mathcal{H}_2)$ and its expectation value,
\begin{equation}
<A>=Tr_{_{\mathcal{H}_2}}(T(\rho)A)=Tr_{_{\mathcal{H}_2}}Tr_{_{\mathcal{K}_2}}[ U(\rho\otimes\sigma )U^{\dag} (A\otimes I)].
\end{equation}
Replacing $I$ from (\ref{measurment}) in the above equation, we get
\begin{eqnarray}
<A>&=&\sum_{\alpha}Tr_{_{\mathcal{H}_2}}\{Tr_{_{\mathcal{K}_2}}[U(\rho\otimes\sigma )U^{\dag} (A\otimes M_{\alpha})]\}\cr
&=&\sum_{\alpha}Tr_{_{\mathcal{H}_2}}(T_{\alpha}(\rho)A),
\end{eqnarray}
where $T_{\a}:\mathcal{L}(\mathcal{H}_1)\rightarrow \mathcal{L}(\mathcal{H}_2)$ is defined as,
\begin{equation}
T_{\alpha}(\rho):=Tr_{_{\mathcal{K}_2}}[U(\rho\otimes\sigma )U^{\dag} (I\otimes M_{\alpha})].
\end{equation}
Rewriting the expectation value of $A$ in the following way,
\begin{equation}
<A>=\sum_{\alpha}p_{\alpha}Tr[\frac{T_{\alpha}(\rho)}{p_{\alpha}}A],
\end{equation}
we can conclude that $p_{\alpha}=Tr(T_{\alpha}(\rho))$ is the probability of getting
$\alpha$ as the classical result of measurement and the density matrix
$\frac{1}{p_{\alpha}}T_{\alpha}(\rho)$ as selected state of the system after measurement. Therefore by performing the
measurement on the environment the channel is decomposed as,
 \begin{equation}\label{Talpha}
 T=\sum_{\alpha}T_{\alpha}.
 \end{equation}
Let us recall from \cite{QLF} that the most informative measurements on the environment are those
for which the selected output of the channel after the measurement can be described by a single Kraus
operator $t_{\a}:\mathcal{H}_1\rightarrow \mathcal{H}_2$,
\begin{equation}
T_{\alpha}(\rho)=t_{\a}\rho t_{\a}^{\dag}.
\end{equation}
From now on we assume that the initial state of the environment is pure. It has been proved
in \cite{QLF} that when the initial state of the environment is pure, every decomposition of
the channel in form (\ref{Talpha}) can be realized by a measurement on the environment. We also
assume that we can perform the most informative measurement on the environment . Therefore designing
different POVMs on the environment is equivalent to considering different Kraus representations of the
original channel:
\begin{equation}\label{decomp}
T(\rho)=\sum_{\alpha}t_{\alpha}\rho t_{\alpha}^{\dag}, \hskip 1 cm \sum_{\alpha}t_{\alpha}^{\dag}t_{\alpha}=I.
\end{equation}
To perform the correction we design
a recovery channel $R_{\alpha}:\mathcal{L}(\mathcal{H}_2)\rightarrow \mathcal{L}(\mathcal{H}_1)$ that depends on the measurement outcomes.
The state of the system after performing the correction is $\frac{1}{p_{\a}}R_{\a}(T_{\a}(\rho))$
with probability $p_{\a}$ and the overall channel $T_{corr}:\mathcal{L}(\mathcal{H}_1)\rightarrow \mathcal{L}(\mathcal{H}_1)$
is given by
\begin{equation}
T_{corr}=\sum_{\a}R_{\a}\circ T_{\a},
\end{equation}
where $\circ$ means composition of the maps. The closer the $T_{corr}$ is to the \textbf{id} map, the more
successful is the scheme to recover quantum information. To quantify the performance of the correction
scheme, we use entanglement fidelity \cite{EntFid} as measure.
For a general channel $\Phi: \mathcal{L}(\mathcal{H})\rightarrow \mathcal{L}(\mathcal{H})$ with Kraus operators $A_k$, the entanglement fidelity
is given by
\begin{equation}
F(\Phi)=\la\Psi| \Phi\otimes I(|\Psi\ra\la\Psi|)|\Psi\ra=\frac{1}{d^2}\sum_k|tr A_k|^2
\end{equation}
where $d=dim \mathcal{H}$ and $|\Psi\ra\in \mathcal{H}\otimes \mathcal{H}$ is a maximally entangled state. We are interested in $F(T_{corr})$, the entanglement fidelity of the corrected map. Given  $T(\rho)$
in (\ref{decomp})
and using the Kraus representation of the recovery channel $R_{\a}: \mathcal{L}(\mathcal{H}_2)\rightarrow \mathcal{L}(\mathcal{H}_1)$:
\begin{equation}
R_{\a}(\rho')=\sum_{\b}r_{\b}^{\a}\rho'r_{\b}^{\a\dag},\hskip 1 cm \sum_{\b}r_{\b}^{\a\dag}r_{\b}^{\a}=I,
\end{equation}
the entanglement fidelity of the corrected channel becomes,
\begin{equation}\label{13}
F(T_{corr})=\frac{1}{d^2}\sum_{\a,\b}|tr(r_{\b}^{(\a)}t_{\a})|^2.
\end{equation}
Entanglement fidelity reaches identity and quantum information can
receive complete correction if and only if for all $\a$,  $t_{\a} t_{\a}^{\dag}=\tau_{\a} I$ with
$\tau_{\a}\geq 0$ and $\sum_{\a}\tau_{\a}=1$ \cite{QLF}. In particular, when $dim \mathcal{H}_1=dim \mathcal{H}_2$,
channels with Kraus operators proportional to unitary are completely correctable. In
this case the optimum measurement on the environment is the one corresponding to the unitary decomposition of the map
and the correction scheme in very clear because the evolution of the system is reversible. \\
\\
For those cases where quantum information can not be restored completely, the correction schemes which
give the best information preserving are known. More precisely,
using the Cauchy-Schwarz inequality, it has been shown \cite{QLF} that for every family of
recovery channels $R_{\a}:\mathcal{L}(\mathcal{H}_2)\rightarrow \mathcal{L}(\mathcal{H}_1)$, the entanglement fidelity $F(T_{corr})$
is such that
\begin{equation}\label{upper}
F(T_{corr})\leq\frac{1}{d^2}\sum_{\a}(tr|t_{\a}|)^2,
\end{equation}
where $|t|=\sqrt{t^{\dag}t}$ is the modulus of $t$. Moreover the recovery scheme through which the maximum
can be attained, is also known \cite{QLF}. Therefore even if quantum information can not
be restored completely, the correction giving the highest value of entanglement
fidelity can be applied. It is important to notice that these results are obtained for
a given measurement or equivalently for a given Kraus representation of the channel and leave
the question of the optimal measurement on the environment unanswered.\\
\\
To find the optimal measurement on the environment, we need to maximize the entanglement fidelity
over all possible Kraus decompositions of the channel. However, to make the problem tractable,
we restrict our attention to
Kraus representations with the same number of Kraus operators as the canonical representation.
To be more specific, we denote the upper
bound in equation (\ref{upper}) by $\tilde{F}(T_{corr})$:
\begin{equation}\label{Ftilde}
\tilde{F}(T_{corr})=\frac{1}{d^2}\sum_{\a}(tr|t_{\a}|)^2
\end{equation}
and maximize it over all possible Kraus representations
with the same number $N$ of Kraus operators as the canonical one.
We start working from the canonical representation
of a given map:
\begin{equation}
T(\rho)=\sum_{k=0}^{N-1} C_k\rho C_k^{\dag}
\end{equation}
with Kraus operators satisfying $tr(C_kC_{k'}^{\dag})=c_k\delta_{k,k'}$, $c_k\in \mathbb{R}^+$. Afterwards, considering a
general
$N$ dimensional
unitary operator $V$, we construct a new set of Kraus operators $\{B_k\}$
for the channel
\begin{equation}
B_k=\sum_{l=0}^{N-1}V_{k,l}C_l,
\end{equation}
and maximize the entanglement fidelity given by the new set of Kraus operators
over the parameters of the unitary matrix.\\
\\
It is important to notice that instead of considering a general unitary operator
in $U(N)$ it is sufficient to perform the maximization over the parameters in $SU(N)$.
This is because any operator $V\in U(N)$ can be written as $V=e^{i\phi}U$
with $U\in SU(N)$ and $\phi\in \mathbb{R}$.
%Consider
%a general unitary operator $U\in SU(N)$ then $V\in U(N)$ can
%simply be constructed by multiplication of an arbitrary phase $V=e^{i\phi}U$.
Constructing the new set of Kraus operators by means of $V$, we get:
\begin{equation}
B_k=\sum_{l=0}^{N-1}V_{k,l}C_l=e^{i\phi}\sum_{l=0}^{N-1}U_{k,l}C_l
\end{equation}
However only the absolute values of $B_k$s play a role in the entanglement fidelity,
\begin{equation}
|B_k|=\sqrt{B_k^{\dag}B_k}=(\sum_{l,l'}U_{k,l}^*U_{k,l'}C_l^{\dag}C_{l'})^{\frac{1}{2}}.
\end{equation}
Thus, the entanglement fidelity does not depend on the phase $e^{i\phi}$ and
therefore we can restrict our attention to Kraus representations obtained by transformations
in $SU(N)$ without loss of generality.\\
\\
In the following Sections we find the optimum measurement
to perform correction on amplitude damping channel. This is a nontrivial case because this
map is not a random unitary map. Therefore, it is not included in the category of
completely correctable maps discussed after equation (\ref{13}), with trivial optimum measurement
and correction scheme.
\section{Amplitude damping channel}\label{AD-Channel}
The general behavior of processes with energy dissipation to the environment is
well characterized by the amplitude damping channel. The gradual dissipation of energy or
amplitude damping channel is described by the interaction between the system and the
environment, modeled by harmonic oscillators, through the unitary operator:
$$U=e^{-i\chi(ab^{\dag}+a^{\dag}b)},$$
where $\chi$ is proportional to the coupling constant between system and environment. The operators $a$ and $a^{\dag}$ (resp $b$ and $b^{\dag}$) denote annihilation and creation operators of the system (resp environment). By truncating the Fock basis of bosonic mode of the system to length
 $d$, we describe a qudit system. The Kraus
operators, $C_{m}=\la m_b|U|0_b\ra$, for this d-dimensional amplitude damping channel are given by
\begin{equation}\label{KO}
 C_m=\sum_{n=m}^{d-1}\sqrt{C(n,m)
 (1-p)^{n-m}p^m}|n-m\ra\la n|, \hskip 0.5 cm m=0\cdots d-1,
\end{equation}
where $p=\sin^2\chi$ is the probability of loosing a single quantum of energy,
$|n\ra$ are number states and $C(n,m)=\left(\begin{array}{c}
 n\cr
 m\end{array}\right)$ is the binomial coefficient. It is straightforward
checking that this is the canonical representation
$$tr(C^{\dag}_mC_{m'})=c_m\delta_{m,m'}, \hskip 1 cm c_m=\sum_{n}C(n,m)(1-p)^{n-m}p^m.$$
Using the definition of the Kraus operators in (\ref{KO}), it is clear that $C_m^{\dag}C_m$ is
diagonal:
\begin{equation}\label{AdagA}
 C_m^{\dag}C_m=\sum_{n=m}^{d-1}C(n,m)(1-p)^{n-m}p^{m}|n\ra\la n|.
\end{equation}
Since amplitude
From equations (\ref{AdagA}) and (\ref{Ftilde}), it is easy to find that the highest value
of the entanglement fidelity for the canonical representation of the amplitude damping channel is,
\begin{eqnarray}
\tilde{F}_c(T_{corr})&=&\frac{1}{d^2}\sum_{m=0}^{d-1}(tr|C_m|)^2\cr
&=&\frac{1}{d^2}\sum_{m=0}^{d-1}\left(\sum_{n=0}^{d-1}
\sqrt{C(n,m)(1-p)^{n-m}p^{m}}\right)^2,
\end{eqnarray}
where the under script $c$ stands for canonical. For systems with two dimensional Hilbert space, the entanglement fidelity for canonical Kraus
representation $\tilde{F}_c(T_{corr})$
is given by,
\begin{equation}\label{F2}
 \tilde{F}_{c}(T_{corr})=\frac{1+\sqrt{1-p}}{2}.
\end{equation}
To see whether or not there is any advantage in using the canonical decomposition, we
start from the canonical representation of the map:
\begin{eqnarray}
C_0&=&|0\ra\la 0|+\sqrt{1-p}|1\ra\la 1|,\cr
C_1&=&\sqrt{p}|0\ra\la 1|.
\end{eqnarray}
considering the general two dimensional special unitary operator
\begin{equation}
 U=\left(\begin{array}{cc}
          \alpha&\beta\cr
         -\overline{\beta}&\overline{\alpha}
         \end{array}\right)\hskip 1cm |\a|^2+|\b|^2=1,
\end{equation}
we introduce a general set of Kraus operators $\{B_k\}$ as follows:
\begin{equation}
B_0=\left(\begin{array}{cc}
          \alpha&\beta\sqrt{p}\cr
         0&{\alpha}\sqrt{1-p}
         \end{array}\right),
\hskip 1 cm
B_1=\left(\begin{array}{cc}
          -\overline{\beta}&\overline{\alpha}\sqrt{p}\cr
         0&-\overline{\beta}\sqrt{1-p}
         \end{array}\right).
\end{equation}
For this set of Kraus operators, the entanglement fidelity is given by
\begin{equation}\label{FB2}
 \tilde{F}_B(T_{corr})=\frac{1}{4}\sum_{m=0}^1tr^2(|B_m|)=\frac{1+\sqrt{1-p}}{2}.
\end{equation}
Comparing equations (\ref{F2}) and (\ref{FB2}), we conclude that all
representations with two Kraus operators give the same entanglement fidelity as the canonical one.\\
\\
To answer the question of whether or not the same result is valid in higher dimensions, we face the problem of diagonalizing  $d$-dimensional matrices $B_k^{\dag}B_k$ to compute
the entanglement fidelity. Therefore in the next Section we study the amplitude damping channel
in three dimension. This will give us some insights on possible advantages (or disadvantages) of
using the canonical representation.
\section{ Amplitude damping channel for qutrits}\label{AD-3}
For three dimensional Hilbert space the canonical decomposition of the
channel is described by the following Kraus operators:
\begin{eqnarray}
 C_0&=&|0\ra\la 0|+\sqrt{1-p}|1\ra\la 1|+(1-p)|2\ra\la2|,\cr
 C_1&=&\sqrt{p}|0\ra\la 1|+\sqrt{2p(1-p)}|1\ra\la 2|,\cr
 C_2&=&p|0\ra\la 2|.
\end{eqnarray}
The entanglement fidelity corresponding to the canonical decomposition is
\begin{equation}\label{Fc}
 \tilde{F}_c(T_{corr})=\frac{1}{9}[(2-p+\sqrt{1-p})^2+(\sqrt{p}+\sqrt{2p(1-p)})^2+p^2].
\end{equation}
As explained in Section \ref{QF}, in order to find the optimal measurement we maximize the entanglement
fidelity over all possible Kraus representations with $N=3$ number of Kraus operators.
To construct the general Kraus
representation from the canonical one, we can restrict ourselves to the unitary operators in $SU(3)$
without loss of generality. As starting working hypothesis, we consider the following two
subgroups of $SU(3)$, $G_1$ and $G_2$, leading to equal or higher entanglement fidelity than
the one given by canonical Kraus representation. We show that the decomposition giving the highest
value of entanglement fidelity can be constructed by means of unitary transformations in  $G_2$.\\
\subsection{Equi-canonical class}
The first subgroup $G_1$ that we study is defined as,
\begin{equation}
G_1:=\{U_1=\left(\begin{array}{ccc}
          1&0&0\cr
          0&\alpha&\beta\cr
          0&-\overline{\beta}&\overline{\alpha}
         \end{array}\right)| \hskip 5mm |\alpha|^2+|\beta|^2=1
         \}.
\end{equation}
The general Kraus operators
constructed in terms of  the elements of this subgroup are given by,
\begin{eqnarray}\label{c1}
B_0&=&C_0,\cr
B_1&=&\a C_1+\b C_2,\cr
B_2&=&-\bar{\b} C_1+\bar{\a} C_2.
\end{eqnarray}
To calculate the entanglement fidelity $\tilde{F}_B(T_{corr})$ for this class of Kraus operators,
$$\tilde{F}_B(T_{corr})=\frac{1}{9}\sum_{k=0}^2 (tr|B_k|)^2,$$
we
need to compute the eigenvalues of the $B_k^{\dag}B_k$s. It turns out that
the non vanishing eigenvalues of $B_1^{\dag}B_1$ are given by
\begin{equation}
 \lambda_1,\lambda'_1=\frac{a\pm\sqrt{a^2-b^2}}{2},
\end{equation}
with
\begin{equation}
 a=p^2+3p(1-p)|\alpha|^2,\hskip 5 mm
 b= 2p|\alpha|^2\sqrt{2(1-p)}.
\end{equation}
The non vanishing eigenvalues of $B_2^{\dag}B_2$ are,
\begin{equation}
 \lambda_2,\lambda'_2=\frac{a'\pm\sqrt{a'^2-b'^2}}{2},
\end{equation}
with
\begin{equation}
 a'=p^2+3p(1-p)|\beta|^2,\hskip 5 mm
 b'=2p|\beta|^2\sqrt{2(1-p)}.
\end{equation}
Therefore the entanglement fidelity is given by
\begin{eqnarray}\label{36}
 \tilde{F}_B(T_{corr})&=&\frac{1}{9}[(2-p+\sqrt{1-p})^2+a+a'+b+b']\cr
&=&\frac{1}{9}[(2-p+\sqrt{1-p})^2+\left(\sqrt{p}+\sqrt{2p(1-p)}\right)^2+p^2].
\end{eqnarray}
Comparing (\ref{36}) with (\ref{Fc}), we see that for all the new
set of Kraus operators ${B_k}$ in (\ref{c1}),  the entanglement fidelity $\tilde{F}_B(T_{corr})$ in equation (\ref{36}) equals the entanglement fidelity for the canonical Kraus representation,
\begin{equation}
\tilde{F}_B(T_{corr})=\tilde{F}_c(T_{corr}).
\end{equation}
This means there is no advantage in using canonical representation over the
class of Kraus decomposition in (\ref{c1}).\\
\\
As a side remark, we notice that other sets of Kraus operators
$\{B'_k\}$ leading to the entanglement fidelity $\widetilde{F}_B(T_{corr})$ in (\ref{36})
can be constructed. Introduce the two $SU(3)$-operators $g_1$ and $g_2$ as follows
\begin{equation}\label{g1g2}
g_1=\left(\begin{array}{ccc}
0&0&1\cr
1&0&0\cr
0&1&0\end{array}\right)\hskip 1cm
g_2=\left(\begin{array}{ccc}
0&1&0\cr
0&0&1\cr
1&0&0\end{array}\right)
\end{equation}
and consider the new set of Kraus operators $\{B'_k\}$ given by
\begin{equation}\label{Bp}
B'_k=\sum_l (g_{j})_{k,l}B_l \hskip 1cm j=1,2.
\end{equation}
From (\ref{Bp}), we notice that the only effect of the $g$s on the vector $B\equiv(B_0,B_1,B_2)^t$ is that of cyclically permuting the vector components.\\
\\
The class of unitary operators
giving the Kraus representations $\{B'_k\}$ from the canonical ones are simply the following left cosets
of $G_1$:
\begin{eqnarray}
g_1G_1&:=&\{\left(\begin{array}{ccc}
0&-\bar{\b}&\bar{\a}\cr
1&0&0\cr
0&\a&\b\end{array}\right)|\hskip 5mm |\a|^2+|\b|^2=1
\},\cr\cr
g_2G_1&:=&\{\left(\begin{array}{ccc}
0&\a&\b\cr
0&-\bar{\b}&\bar{\a}\cr
1&0&0\end{array}\right)|\hskip 5mm |\a|^2+|\b|^2=1
\},
\end{eqnarray}
Therefore any Kraus representation obtained from the canonical one via transformations in
$$G_1\bigcup g_1G_1 \bigcup g_2G_1$$
leads to the same entanglement fidelity $\tilde{F}_c(T_{corr})$ in (\ref{Fc}).
\subsection{Super-canonical class}
In the previous subSection we showed there is a large class of Kraus representations
leading to the same entanglement fidelity that can be achieved by the canonical
representation. A more interesting question is whether or not we can
design a measurement on the environment that gives enatanglement fidelity values higher than the
canonical one. To answer this question, we consider a new subgroup $G_2$ of $SU(3)$,
\begin{equation}
G_2:=\{U_2=\left(\begin{array}{ccc}
\gamma&0&\delta\cr
0&1&0\cr
-\bar{\delta}&0&\bar{\gamma}
\end{array}\right)| \hskip 5mm |\gamma|^2+|\delta|^2=1\}.
\end{equation}
In terms of $U_2$, the new set of Kraus operators becomes
\begin{eqnarray}
D_0&=&\gamma C_0+\delta C_2,\cr
D_1&=&C_1,\cr
D_2&=&-\bar{\delta} C_0+\bar{\gamma} C_2.
\end{eqnarray}
Following the line of reasoning presented in the previous subSection, we obtain
\begin{equation}\label{tD0}
tr|D_0|=\left(\frac{g+\sqrt{g^2-h^2}}{2}\right)^{\frac{1}{2}}+\left(\frac{g-\sqrt{g^2-h^2}}{2}\right)^{\frac{1}{2}}+\sqrt{1-p}|\gamma|
\end{equation}
with,
\begin{eqnarray}
g&=&p^2+2(1-p)|\gamma|^2,\cr
h&=&2(1-p)|\gamma|^2.
\end{eqnarray}
Similarly
\begin{equation}\label{tD2}
tr|D_2|=\left(\frac{k+\sqrt{k^2-l^2}}{2}\right)^{\frac{1}{2}}+\left(\frac{k-\sqrt{k^2-l^2}}{2}\right)^{\frac{1}{2}}+\sqrt{1-p}|\delta|
\end{equation}
where
\begin{eqnarray}
k&=&p^2+2(1-p)|\delta|^2,\cr
l&=&2(1-p)|\delta|^2.
\end{eqnarray}
Therefore the entanglement fidelity obtained from the new set of Kraus representation becomes (Appendix A)
\begin{equation}\label{FdFcOmega}
\tilde{F}_D(T_{corr})=\tilde{F}_c(T_{corr})+\frac{2\sqrt{1-p}}{9}\Omega,
\end{equation}
with
\begin{equation}\label{Omega}
\Omega=|\gamma|\sqrt{g+h}+|\delta|\sqrt{k+l}-(2-p).
\end{equation}
Since $\Omega$ is strictly positive (Appendix A), it follows that
\begin{equation}
\tilde{F}_D(T_{corr})>\tilde{F}_c(T_{corr}).
\end{equation}
%%%%%%%%%%%%%%%%%%%%%%%%%%%%%%%%%%%%%%%
\begin{figure}[t]
\centering
\includegraphics[height=0.3\textheight]{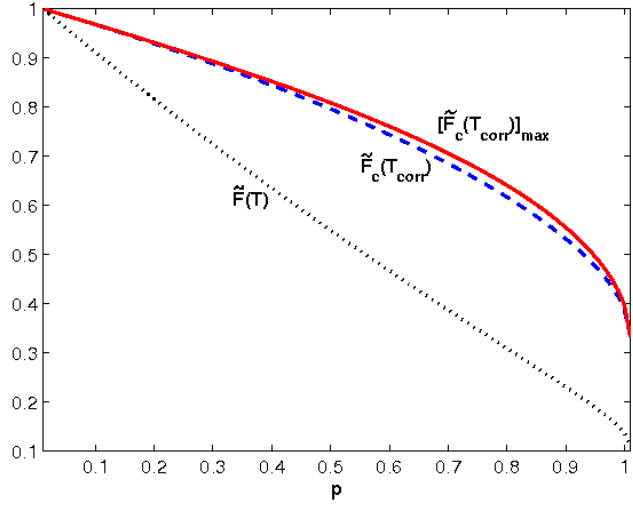}
\caption{(color online) Entanglement fidelity for the equi-canonical class (dash line) and the maximum entanglement fidelity (solid line)
found in the super-canonical class of Kraus representations. Dot line shows entanglement fidelity without
performing any correction.} \label{FcanFmax}
\end{figure}
%%%%%%%%%%%%%%%%%%%%%%%%%%%%%%%%%%%%%%%%%%%%%%%%
We conclude that all the Kraus decompositions constructed using the elements of group $G_2$ lead to entanglement
fidelity values higher than the one obtained by means of the canonical representation. Using the same arguments of the previous
subSection, we can find a larger class of unitary transformations giving rise to Kraus representations with higher entanglement fidelity than the canonical one.
Such transformations belong to the following left cosets of $G_2$,
\begin{eqnarray}
g_1G_2&:=&\{\left(\begin{array}{ccc}
-\bar{\delta}&0&\bar{\gamma}\cr
\gamma&0&\delta\cr
0&1&0\end{array}\right)|\hskip 5mm |\gamma|^2+|\delta|^2=1
\},\cr\cr
g_2G_2&:=&\{\left(\begin{array}{ccc}
0&1&0\cr
-\bar{\delta}&0&\bar{\gamma}\cr
\gamma&0&\delta\end{array}\right)|\hskip 5mm |\gamma|^2+|\delta|^2=1
\},
\end{eqnarray}
where $g_1$ and $g_2$ are given in equation (\ref{g1g2}).
Therefore the operators in the set
\begin{equation}\label{set}
G_2\bigcup g_1G_2 \bigcup g_2G_2
\end{equation}
give rise to decompositions that work better than the canonical one. We can find the best decomposition
in this class by maximizing $\Omega$ in equation (\ref{Omega}) over the parameters defining the transformations in
class (\ref{set}). It follows that the maximum is achieved for $|\gamma|=|\delta|=\frac{1}{\sqrt{2}}$:
\begin{equation}
\Omega_{max}=\sqrt{2+2(1-p)^2}-(2-p).
\end{equation}
Thus, the maximum entanglement fidelity in this class becomes
\begin{equation}\label{max}
[\tilde{F}_D(T_{corr})]_{max}=\frac{1}{9}\left[5-2p+2p\sqrt{2(1-p)}+2\sqrt{2(1-p)(2-2p+p^2)}\right],
\end{equation}
and can be achieved by decompositions arising from the canonical one by means of the following
special unitary transformations:
\begin{equation}
U_{G_2}=\left(\begin{array}{ccc}
\frac{e^{i\theta}}{\sqrt{2}}&0&\frac{e^{i\phi}}{\sqrt{2}}\cr
0&1&0\cr
-\frac{e^{-i\phi}}{\sqrt{2}}&0&\frac{e^{-i\theta}}{\sqrt{2}}
\end{array}\right)\hskip 5mm
U_{g_1G_2}=\left(\begin{array}{ccc}
-\frac{e^{-i\phi}}{\sqrt{2}}&0&\frac{e^{-i\theta}}{\sqrt{2}}\cr
\frac{e^{i\theta}}{\sqrt{2}}&0&\frac{e^{i\phi}}{\sqrt{2}}\cr
0&1&0
\end{array}\right)\hskip 5mm
U_{g_2G_2}=\left(\begin{array}{ccc}
0&1&0\cr
-\frac{e^{-i\phi}}{\sqrt{2}}&0&\frac{e^{-i\theta}}{\sqrt{2}}\cr
\frac{e^{i\theta}}{\sqrt{2}}&0&\frac{e^{i\phi}}{\sqrt{2}}
\end{array}\right)\hskip 5mm
\end{equation}
Figure \ref{FcanFmax} shows the entanglement fidelity for the case that no correction has been
perfomed (dot line), for the equi-canonical class (dash line) and the maximum entanglement
fidelity (solid line) that can be attained in the super-canonical class.
\subsection{Maximum entanglement fidelity}
It is impractical
to analytically prove that the entanglement fidelity in equation (\ref{max}) is
the global maximum not just the local maximum in the super-canonical
class. However, we are able to show this numerically. We generate $n=10^5$ sets of random
unitary operators using the fact that an arbitrary operator in $SU(3)$ is generated by Gell-Mann
matrices, $\{\Lambda_j\}$ \cite{Gell-Mann}:
\begin{equation}
 U=\exp[-i\sum_{j=1}^8 a_j \Lambda_j],
\end{equation}
where $a_i$ are real coefficients. The chosen cardinality $n$ of the randomly generated
distinct Kraus representations is the smallest positive integer number necessary to obtain convergence
to the numerically found global maximum. Any other randomly generated set of Kraus representations with
higher cardinality $m>n$ converges to the same global maximum. Our numerical analysis implies that such global maximum coincides with the analytical expression in equation
(\ref{max}). Thus, we conclude that $[F_D(T_{corr})]_{max}$ in (\ref{max}) is indeed the global maximum.\\
\section{Discussion}\label{diss}
The main emphasis of this paper is to reclaim quantum information lost to the environment
surrounding the system. Although the evolution of the system is irreversible, but having access
to the environment enables us to restore quantum features of the system. The main idea
is to perform appropriate correction on the system regarding
the classical results obtained by measuring the environment.This scheme is an
example of feed back control. Here
we have addressed the important question of what kind of measurement provides us with the highest
recovery performance.\\
\\
To  answer this question, we have applied quantum feed back control scheme to amplitude
damping channel describing the loss of
information to the environment. After describing the amplitude damping channel in
arbitrary dimension we have concentrated on two dimensional systems. We showed that all measurements
corresponding to Kraus representations of the map with two Kraus operators (measurements with
two outcomes) give the same
value for the entanglement fidelity. It implies that for the amplitude damping channel in
$d=2$, the reduction of decoherence using the
quantum feedback control scheme
does not depend on the details of the measurement made on the environment.
Our numeric studies show that the same statement is valid even if we consider measurements
with three or four outcomes.\\
\\
Although in two dimensional Hilbert space the performance of quantum feedback control
does not depend on measurement details, the situation becomes different when we increase
the dimension to three. By studying the three dimensional amplitude damping channel, we
have showed there is a class of measurements or class of Kraus representations of the
amplitude damping channel which perform as good as canonical representation. We named it
equi-canonical class and analytically found the entanglement fidelity
that can be attained by performing such measurements. Interestingly, we introduced another
class of Kraus representations, the super-canonical class, which leads to an entanglement fidelity
higher than the one obtained in the equi-canonical class. We analytically found the maximum
entanglement fidelity in the super-canonical class and the Kraus representations by which this maximum can
be attained. By means of numeric techniques, we discovered that the maximum entanglement
fidelity we have found is not only the maximum
in this class but also the global maximum over all possibilities when considering the most general
Kraus representations with three Kraus operators. Furthermore
our numeric studies shows that the same statement is true even if we
consider measurement with four outcomes. Indeed, motivated by numerical evidences, we feel like conjecturing that this might be true for any number of Kraus operators (measurement outcomes).\\
\\
{\Large{\textbf{Acknowledgement}}}\\
\\
L.M. and S.M. would like to thank G. Chiribella, G.M. D'Ariano and P. Perinotti for useful discussions in the early stage of this work.
The authors acknowledge financial support by the European Commission,
under the FET-Open grant agreements HIP (number FP7-ICT-221889)
and CORNER (number FP7-ICT-213681).\\
\\
{\Large{\textbf{Appendix A}}}\\
\\
In what follows, we justify the positivity of $\Omega$ in (\ref{Omega}).
Recall that $\tilde{F}_D(T_{corr})$ is given by,
$$\tilde{F}_D(T_{corr})=\frac{1}{9}\sum_{k=0}^2(tr|D_k|)^2,$$
with $D_1=C_1$. Therefore $\tilde{F}_D(T_{corr})$ becomes,
\begin{equation}
\tilde{F}_D(T_{corr})=\frac{1}{9}[(tr|C_1|)^2+(tr|D_0|)^2+(tr|D_2|)^2].
\end{equation}
Replacing $tr|D_0|$ and $tr|D_2|$ from equations (\ref{tD0}) and (\ref{tD2}) into the
above equation and using the following identity
\begin{equation}\label{key}
|\alpha|\sqrt{a}+|\beta|\sqrt{b}=\sqrt{|\alpha|^2a+|\beta|^2b+2|\alpha||\beta|\sqrt{ab}},
\end{equation}
we obtain
\begin{equation}
\tilde{F}_D(T_{corr})=\frac{1}{9}[tr^2(|C_1|)+2p^2+5(1-p)+2\sqrt{1-p}(|\gamma|\sqrt{g+h}+|\delta|\sqrt{k+l})].
\end{equation}
Using equation (\ref{Fc}), we can rewrite the above equation to get equation (\ref{FdFcOmega}),
$$
\tilde{F}_D(T_{corr})=\tilde{F}_c(T_{corr})+\frac{2\sqrt{1-p}}{9}\Omega
$$
with
$$\Omega=|\gamma|\sqrt{g+h}+|\delta|\sqrt{k+l}-(2-p).$$
To prove that $\Omega >0$, we first use (\ref{key}) to rewrite $\Omega$ as follows,
\begin{eqnarray}\label{akhar}
\Omega&=&\sqrt{|\gamma|^2(g+h)+|\delta|^2(k+l)+2|\gamma||\delta|\sqrt{(g+h)(k+l)}}-(2-p)\cr
&=&\sqrt{(2-p)^2-8(1-p)|\gamma|^2|\delta|^2+2|\gamma||\delta|\sqrt{p^4+4p^2(1-p)+16(1-p)^2|\gamma|^2|\delta|^2}}-(2-p)\cr
\end{eqnarray}
Since the following inequality holds,
\begin{equation}
2|\gamma||\delta|\sqrt{p^4+4p^2(1-p)+16(1-p)^2|\gamma|^2|\delta|^2}>8(1-p)|\gamma|^2|\delta|^2,
\end{equation}
we conclude that $\Omega>0$.\\

\end{document}